# A Course on Controllers

**Bill Verplank, Craig Sapp, Max Mathews**
Center for Computer Research in Music and Acoustics
Department of Music, Stanford University
Stanford, California 94305-8180, USA
http://www-ccrma.stanford.edu/courses/250a

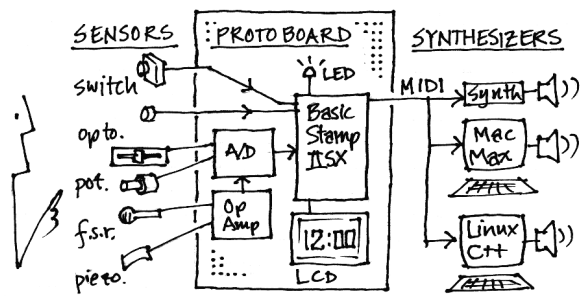

## ABSTRACT
Over the last four years, we have developed a series of lectures, labs and project assignments aimed at introducing enough technology so that students from a mix of disciplines can design and build innovative interface devices.

## INTRODUCTION
In 1996, Perry Cook and Ben Knapp at Princeton and San Jose State received an NSF grant to develop a multi-campus course on the technology of human-computer interaction. Stanford joined in three-way teleconferences for the next three years [1]. In 1999, we did a network conference with just Stanford and Princeton. This year, we abandoned the conferencing and taught a Stanford-only course.

Over the years, we have come to focus less on theory and more on practical skills leading to a four-week project: designing and building a working controller.

## Theme
The basic theme of the course is the choice of buttons or handles. The students found examples and did sketches for in-class discussion.

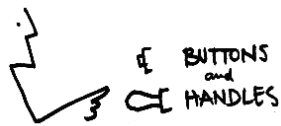

We agreed that the most successful interfaces provide a spectrum: from discrete to continuous control. Our belief is that innovative styles of interaction for a variety of products and systems will result from exploring this range.

## LABS
We prepared six laboratory exercises covering discrete and continuous sensors, electronics and some simple music programming. To connect sensors to computers, we designed a simple prototyping board and built enough copies so each pair of students could have their own board.

### Prototyping Board
The board was a "classic" wood breadboard. On the board are mounted two or three commercial solder-less breadboards which have holes at 0.1 inch spacing. The breadboards allow the rapid construction and destruction of circuits by plugging in integrated circuit chips, resistors etc, and connecting elements with precut jumpers. On the board we mounted a 9-pin D connector as interface with a computer serial port and two 5-pin female DIN connectors, one for MIDI output and one for a standard commercial power supply input. The power supply provides three voltages, 5 volts for the microprocessor, and plus and minus 12 volts for standard op amps. The same power supply drives the Radio-baton [6].

Although the prototyping boards can be used in other ways, most boards contain a Parallax Basic Stamp IISX microprocessor module [2] and a Maxim Max 1270 8-channel 12-bit A-D converter. The Basic Stamp module includes an EEPROM memory which can be programmed from a PC with variant of Basic. The program is downloaded and written into the EEPROM through the serial port. The program remains in the EEPROM until it is rewritten, even if power is disconnected.

The Basic Stamp communicates via 16 I-O pins. Four pins are dedicated to the A-D converter, one pin is dedicated to MIDI output. The remaining pins can be used arbitrarily. We provide a program which sequentially reads the 8 A-D channels and 7 I-O pins which have 0 or 5 volt inputs from buttons (switches). The program encodes this information as 9 standard 3-byte MIDI commands and sends these commands via the MIDI DIN connector to any device-- computer or a synthesizer-- that can read MIDI. With this program, only the most significant 7 bits of the 12-bit A-D outputs are transmitted since most MIDI commands are limited to 7-bit data. However, it is not hard to alter the program to send all 12 bits. Craig Sapp has put much of what we use on line. [3]

MIDI was chosen as the communication media because most of our students are music majors and MIDI is almost universally used in the domains of electronic music. Almost any PC or MAC – and almost any operating system – MAC, Windows, or Linux – can send and receive standard MIDI commands.

Most sensors communicate via signals represented as voltages, but these signals often need amplification or attenuation or filtering to be appropriate for the inputs of the A-D converter. It is easy to plug op amps into the board for this purpose. We provide the students with basic "cookbook" training sufficient to design and build amplifiers, mixers, low-pass filters to remove noise, bandpass filters and enveloping filters. .

The Basic Stamp and A-D converter running at maximum speed transmits a complete cycle of 8 MIDI encoded analog signals and one MIDI encoded 7-button signal in 36ms. Although not outstandingly fast, this is a useful speed for many human-computer interactions. cycle requires 27 bytes of MIDI code. A MIDI channel can send about 3 bytes per millisecond so 27 bytes requires about 9ms. Thus, the Basic stamp uses about 1/4 of the capacity of the MIDI channel. The Scenix microcontroller on the Basic Stamp could run much faster with an assembly language program rather than with the interpreted Basic commands which we now use.

The cost of the principal components in a prototyping board is:

| | |
|---|---|
| Basic stamp | $70 |
| A-D converter | $20 |
| Power supply | $35 |
| Solder less Breadboards | $15 |
| Total | $140 |

**Microprocessor Programming**

The first two labs consisted of designing and programming a simple count-down timer on an LCD display. The design problem introduced the concepts of "modes" and "mappings", the implementation involved writing simple Basic language routines for reading buttons, counting and timing and sending serial display messages to the LCD.

In past years, we had found too late, that the projects all benefited from use of microprocessors so this year we started early. By project time, every team had sufficient prototyping skills.

**Sensors and Music**

Two labs introduced continuous sensors (linear and rotary potentiometers, force-sensitive resistors, piezo-film, sonar, photo-resistors), signal conditioning (amplifying, offsetting and filtering with an operational amplifier), A/D conversion, and MIDI output to a PC running a simple square-wave synthesizer written in C++ on Linux.

For music, building an electronic music controller is only the first step; you must then figure out how your instrument will make music. A C-based programming environment, called Improv, for Windows and Unix was developed for courses that teach algorithmic composition programming with the Radio Baton 3D position controller and musical keyboards as input [4].

Simple electronic instruments may not need to be linked to a computer and can send performance messages directly to a synthesizer via MIDI; however, with computer-based devices, the compositional aspects of music can be intermingled with instrumental performance. With a programming environment such as Improv, or the MAX visual programming environment for Macintosh, students in the subsequent courses address the issues of making music with the instruments they have invented.

**Human Performance**

Two labs were dedicated to measuring human performance on various tasks (e.g. Fitts' Tapping task, Zhai's Steering task [6]) using various input devices: Mathew's Radio Baton [6], linear and rotary potentiometers, Massie's Phantom[7]. With the Phantom, we were able to vary dynamics (mass, damping, friction) with computer-synthesized force-feedback and to compare human performance and preferences.

**PROJECTS**

**Interaction Design Process**

To start the projects we used the following frame-work:

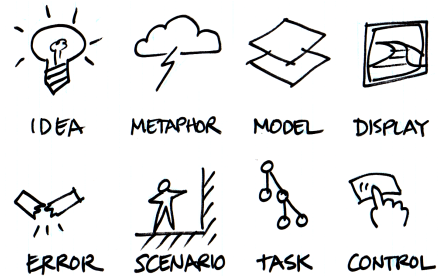

We illustrate the framework with Celine Perrin's project: A two-way "Haptic Pager". The ERROR or annoyance is that cell phones ring in public. Her IDEA is a one-to-one silent and personal link: when holding hands, give a squeeze (METAPHOR). One SCENARIO has Sam at home wondering if Sally is just stuck in a check-out line at the store.

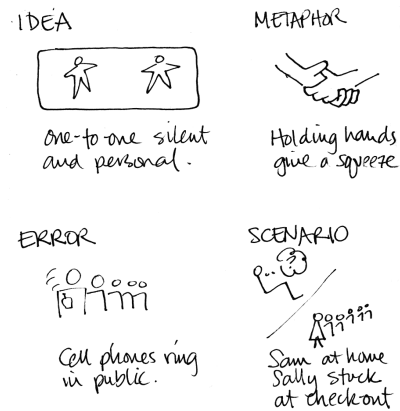

The necessary users' conceptual MODEL is to think of it as 1. A Single Channel where Sam and Sally are directly connected, and 2. Packets which are sent out, and at some time latter, replies are received. The corresponding TASK involves a SET-UP mode where the Sam/Sally link is chosen and then a SQUEEZE mode where haptic messages are exchanged. A proposed DISPLAY shows a list of people and the corresponding CONTROL is selecting with a tap of the stylus. The vibrator DISPLAY might be on a necklace for receiving and the CONTROL would be a squeeze of the necklace for sending.

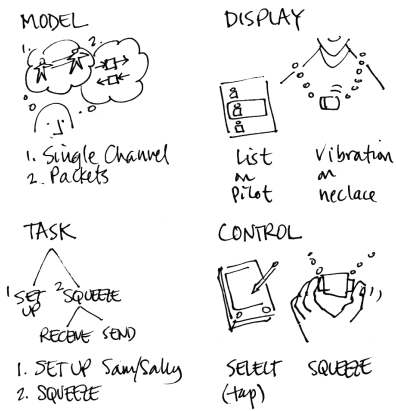

The important thing about the framework is to use it as a check on the balance of approaches from invention to implementation and from overviews to details. It is not intended as a strict ordering of the invention and design process. Most of the projects from the class actually started with some sort of CONTROL idea and only later considered whose problem they might be solving.

**PROJECTS**

There were six projects this year with teams ranging in size from one to four. To our delight they were all able to show working systems where at least some input caused some output. In the next quarter, the students will continue development and refinement.

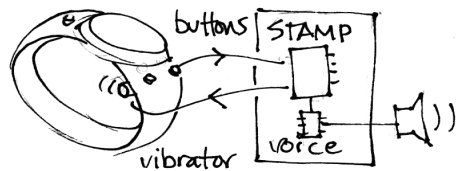

- Celine Pering built her haptic pager (vibrator) into a wrist-band with three buttons. She programmed the Basic Stamp to drive the vibrator with distinct patterns as well as to speak one of three messages from a voice chip. This was the project that showed the most complicated development using just the Stamp; now she can begin to look at the issues of making it two-way.

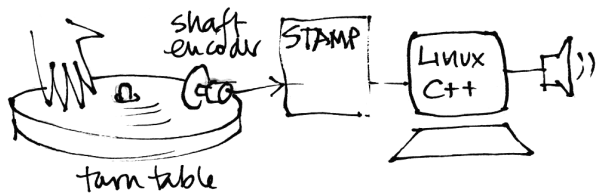

- Keatly Halderman, Daniel Lee, Steve Perella and Simon Reiff connected an optical shaft encoder to a wheel riding on a record on a turntable. The Stamp converted pulse-widths to velocity and sent it to a PC via MIDI where a C++ program played back a recorded sample with that rate. This was the most direct translation of an existing "instrument" into the digital realm; now they can begin to explore the digital advantages, like instant groove selection.

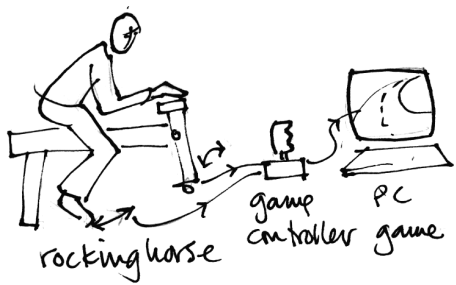

- Pascal Soboll built a big wooden horse for riding and simply sent voltages from two potentiometers through a game stick to a PC-based driving simulation. He will continue to explore such "whole-body" driver controls; this was just his first working system.

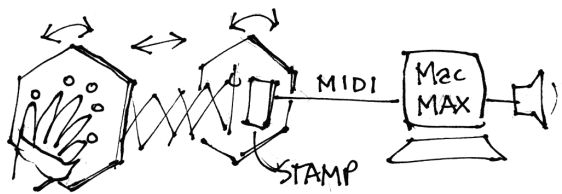

- Michael Gurevich and Stephan Von Muehlen built an accordian-like mechanism with three continuous degrees of freedom plus five buttons on each end. All the signal conditioning is done with an on-board Stamp which sends MIDI to a Macintosh running an FM synthesis patch in Max. This was the most complete project combining an innovative controller with computer music.

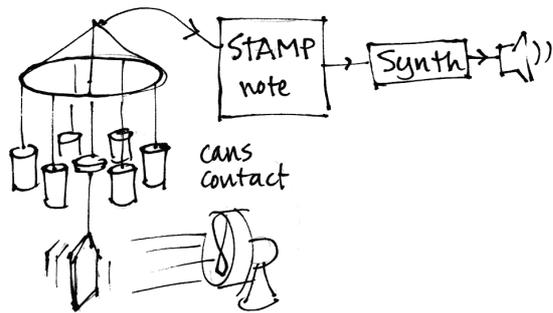

- Zaz Harris and Nathan Schuett built a wind-chime from beer cans. The Stamp senses contacts between the wind-blown clapper and the separate cans and sends off different MIDI messages directly to a commercial synthesizer. Of all the projects, this produced the most surprising and satisfying "musical" results.

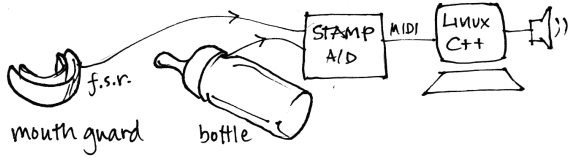

- Molly Norris and Shuli Gilutz explored the use of mouth mounted sensors for a variety of users and uses. Two models were build using force-sensitive-resistor strip that senses contact position: one based on a mouth guard the other on a baby-bottle. This is potentially a rich area of exploration with many applications beyond music; it needs better sensors and some clever application development.

**CONCLUSIONS**

We are always surprised at the creativity of students faced with the challenge of making something that works, but at the same time, we need to provide the time to reflect on what has been learned. While team-work mixes talents and accomplishes more, the students want more time to individually master the technology and design process.

Our goals are innovation and understanding. We believe that the direct engagement in an expressive realm like music can generalize to a wide range of human-machine controllers.

**ACKNOWLEDGMENTS**
Perry Cook and Ben Knapp started it. Terry Winograd and Chris Chafe encouraged it. Shumin Zhai has lectured every year. Other lecturers over the years have been Sile O'Modhrain, Dick Duda, David Jaffe, Matt Gorbet, Bob Adams. Our teaching assistants have been Bill Putnam, Tim Stilson, Craig Sapp, Florian Vogt, Tamara Smyth and Aaron Hipple. Johnathan Berger taught with us this year.